\documentstyle[epsfig,times]{l-aa}

\newcommand{\bgeq}{\begin{equation}}
\newcommand{\edeq}{\end{equation}}
\newcommand{\bgar}{\begin{array}}
\newcommand{\edar}{\end{array}}
\newcommand{\bgea}{\begin{eqnarray}}
\newcommand{\edea}{\end{eqnarray}}
\newcommand{\nnum}{\nonumber}

\newcommand{\chr}{\it\Gamma\/}
\newcommand{\sfrac}[2]{\mbox{$\frac{#1}{#2}$}}

\newcommand{\spar}[2]{\frac{\partial#1}{\partial#2}}
\newcommand{\dpar}[2]{\frac{\partial^{2}#1}{\partial#2^2}}
\begin{document}
\thesaurus{06(02.07.1; 02.09.1; 02.18.8; 08.09.3; 08.18.1)}
\title{Spontaneous symmetry breaking of rapidly rotating stars
in general relativity: influence of the 3D-shift vector}
\author{S.~Bonazzola, J.~Frieben, \and E.~Gourgoulhon }
\offprints{J.~Frieben}
\institute{D\'epartement d'Astrophysique Relativiste et de
  Cosmologie (UPR 176 du CNRS), Observatoire de Paris, \\
   Section de Meudon, F-92195 Meudon Cedex, France \\
  {\em e-mail : frieben@obspm.fr} }
\date{Received 30 June 1997 / Accepted 17 September 1997}
\maketitle
\markboth{S.~Bonazzola et al.: Symmetry breaking of rapidly rotating stars}
{S.~Bonazzola et al.: Symmetry breaking of rapidly rotating stars}
\begin{abstract}
An analytical scheme and a numerical method in order to study the
effects of general relativity on the viscosity driven secular
bar mode instability of rapidly rotating stars are presented.
The approach consists in perturbing an axisymmetric and stationary
configuration and studying its evolution by constructing a series
of triaxial quasi-equilibrium configurations.
These are obtained by solution of an approximate set of field
equations where only the dominant non-axisymmetric terms are taken
into account.
The progress with respect to our former investigation consists in
a higher relativistic order of the non-axisymmetric terms included
into the computation, namely the fully three-dimensional treatment
of the vector part of the space-time metric tensor as opposed to
the scalar part, solely, in the former case.
The scheme is applied to rotating stars built on a polytropic
equation of state and compared to our previous results.
The 3D-vector part turns out to inhibit the symmetry breaking
efficiently. Nevertheless, the bar mode instability is still
possible for an astrophysically relevant mass of
$M_{\rm ns}\!=\!1.4\,M_{\sun}$ when a stiff polytropic equation
of state with an adiabatic index of $\gamma\!=\!2.5$ is employed.
Triaxial neutron stars may be efficient emitters of gravitational
waves and are thus potentially interesting sources for the
forthcoming laser interferometric gravitational wave detectors such
as LIGO, VIRGO and GEO600.
From a numerical point of view, the solution of the three-dimensional
minimal-distortion shift vector equation in spherical coordinates is
an important achievement of our code.
\end{abstract}
\section{Introduction} \label{SEC:INT}
Large scale laser interferometric gravitational detectors such as
LIGO, VIRGO and GEO600 are supposed to become operative in a few
years, and among the possible astrophysical scenarios that of a
rapidly rotating neutron star, deviating progressively from the
initially axisymmetric and stationary configuration towards a
triaxial one, is of particular interest.
In this case, a {\em ``bar shaped''\/} neutron star could become an
efficient source of continuous wave gravitational radiation.
Rapidly rotating stars can break their symmetry, if the ratio
$T/|W|$ of kinetic and potential energy exceeds some critical value
(see Schutz \cite{SCH87} for a review) and the $(l\!=\!2, m\!=\!2)$ bar
mode gets unstable.
This can be the case for a newborn neutron star, having acquired a
sufficiently large amount of rotational kinetic energy during its
formation.
The matter viscosity being low, it may be sensitive to the
{\em Chandrasekhar-Friedman-Schutz\/} instability which is related
to gravitational radiation reaction (Wagoner \cite{WA84}; Lai \& Shapiro
\cite{LS95}).
An alternative mechanism is the {\em viscosity driven secular
instability\/} which is eventually operational in old neutron
stars in close binary systems, being spun up by matter accretion
from a nearby companion to some critical value of $T/|W|$
(Ipser \& Managan \cite{IM84}; Bonazzola et al. \cite{BFG96}).
As concerns rigidly rotating, homogeneous and
self-gravitating fluid bodies, Newtonian theory tells that at
moderate angular velocity, the fluid body is shaped like some
axisymmetric {\em Maclaurin\/} spheroid. At $T/|W|\!=\!0.2738$,
the Maclaurin spheroids become dynamically unstable. However,
at $T/|W|\!=\!0.1375$ exists a bifurcation point towards two
families of triaxial configurations via the above introduced
secular instabilities.
In the case of the viscosity driven instability, kinetic energy is
dissipated whereas angular momentum is conserved. As a consequence,
the fluid body evolves along a sequence close to some
{\em Riemann S\/} ellipsoids towards a {\em Jacobi\/} ellipsoid
which is the configuration with the lowest rotational kinetic energy
for the given angular momentum and mass of the initial Maclaurin
spheroid (Press \& Teukolsky \cite{PT73}).
The transition towards the Jacobi ellipsoid occurs on the
associated {\em viscous\/} time-scale which is much longer than the
dynamical one. For this reason, it is called a {\em secular\/}
instability. In the final state, viscous dissipation has ceased,
and the fluid body rotates rigidly again about its smallest axis
in an inertial frame. It is this viscosity driven secular
instability which we are concerned with in the following.
In the Newtonian compressible case, it was shown by Jeans
(\cite{JE19}, \cite{JE28}), that for a polytropic equation of
state $\gamma\!\ga\!2.2$ is needed to reach the bifurcation
point. If the equation of state is softer, the critical
angular velocity $\Omega_{\rm crit}$ is greater than the mass
shedding limit $\Omega_{\rm K}$, and the bifurcation point is
inaccessible. James (\cite{JA64}) has given a refined value for the
critical adiabatic index of $\gamma_{\rm crit}\!=\!2.238$,
a value which has been recently confirmed by Bonazzola et al.
(\cite{BFG96}) and, independently, by Skinner \& Lindblom (\cite{SL96})
who have found a corresponding value of $\gamma_{\rm crit}\!=\!2.237$.
The result of Bonazzola et al. (\cite{BFG96}) was a by-product of the
investigation of the spontaneous symmetry breaking of fully
relativistic rapidly rotating stars, presented in the same paper.
Until then, all studies of the bifurcation point had been carried
out for Newtonian configurations (Jeans \cite{JE28}; James \cite{JA64};
Ipser \& Managan \cite{IM81}; Hachisu \& Eriguchi \cite{HE82};
Skinner \& Lindblom \cite{SL96})
or at the {\small 1-PN} level of a post-Newtonian analysis,
at most (Chandrasekhar \cite{CH67}; Tsirulev \& Tsvetkov \cite{TT82}).
In view of the highly relativistic character of realistic neutron
star models, this approach is rather motivated by technical
simplification than by physical reasoning.
Our approach consists in perturbing the $(l\!=\!2, m\!=\!2)$ bar mode
of an {\em ``exact''\/} axisymmetric configuration and studying the
growth or the decay of the applied perturbation. We take into
account only the dominant non-axisymmetric terms up to a certain
order. In Bonazzola et al. (\cite{BFG96}), the level of approximation
corresponded to an expansion of the Einstein equations of order
{\small 0-PN} of the three-dimensional terms.
In the present paper, we have included the fully three-dimensional
treatment of the shift vector $N^{i}$, raising the approximation
level to order {\small 1/2-PN}. We recall the basic assumption of
{\em rigid rotation\/} in our approximation (see Bonazzola et al.
(\cite{BFG96}) for a discussion of the astrophysical context), and
the negligibility of gravitational radiation.
Approximation schemes which exploit the near equilibrium of the
gravitational fields, have gained increasing interest in the
context of the binary coalescence problem.
Wilson \& Mathews (\cite{WM89}) were the first ones to propose the
solution of dynamical problems in general relativity by integration
of a reduced set of field equations, which basically corresponds to
the successive solution of initial value problems for the gravitational
fields (York \cite{YO79}) whereas the hydrodynamic part is calculated
exactly under the action of the momentary gravitational fields.
It has been applied in the following to study the
late stages of an inspiralling neutron star binary system (Wilson \&
Mathews \cite{WM95}; Wilson et al. \cite{WMM96}).
Cook et al. (\cite{CST96}), being interested in the initial value
problem of coalescing binaries, have investigated the validity of
Wilson's approach by application to stationary and axisymmetric
neutron star models which, by construction, are perfect equilibrium
systems. They further modified Wilson's approach in a way to
exploit the near equilibrium of the matter fields as well, which
allows to reduce the matter evolution equations to an algebraic first
integral of motion.
The neutron star models computed by means of the simplified field
equations coincide with those based on the full set of Einstein
equations within a few percent.
One of the essential points of the Wilson approach is to adopt a
{\em conformally flat\/} three-metric in order to simplify the field
equations. This {\em ``conformally flat condition''\/} is being
justified by the negligible impact of the radiation part of the
gravitational fields on the matter motion. Indeed, for problems
in spherical symmetry where no radiative component exists, it is
possible to adopt isotropic coordinates without any analytic
approximation.
However, it has to be pointed out, that even for stationary and
axisymmetric systems where no radiative part is present either,
isotropic coordinates are not sufficient to give a complete description
of the problem. A further degree of freedom of the gravitational field
is required which is subject to the dynamical Einstein equations.
The other gravitational fields are then entirely amenable to
determination by means of elliptic equations derived from (1) the
Hamiltonian constraint equation for the conformal factor, (2) the
momentum constraint equation for the shift vector, and (3) the
additional maximal slicing condition which determines the lapse
function. Our model improves on Wilson's scheme in so far as
our approximate equations reduce to the {\em exact\/} Einstein
equations, as long as the configuration remains stationary and
axisymmetric, so that only the three-dimensional perturbation
is treated approximately.

The paper is organized as follows: In Sect.~\ref{SEC:TM} we present
the basic assumptions of our model and derive the field and matter
equations. Sect.~\ref{SEC:NM} introduces the numerical code, and in
Sect.~\ref{SEC:RP} we present the results of the improved scheme
for a polytropic equation of state, including the fully
three-dimensional solution of the shift vector equation, as well
as a comparison with the results of our previous study.
In Sect.~\ref{SEC:CC} we will draw some concluding remarks.
\section{Theoretical model}
\label{SEC:TM}
\subsection{Basic assumptions}
\label{SUBSEC:BA}
The general space-time line element in terms of the quantities
of the (3+1)-formalism of general relativity (see e.g. York
\cite{YO79} for an introduction) is given by
\bgeq \label{EQU:DS2} g_{\mu\nu} \, {\rm d} x^{\mu} {\rm d} x^{\nu} =
   -N^{2} {\rm d} t^{2} + h_{ij}\,({\rm d} x^{i}\!-\!N^{i} {\rm d} t)
   ({\rm d} x^{j}\!-\!N^{j} {\rm d} t) \ ,
\edeq
with the {\em lapse function\/} $N$, the {\em shift vector\/}
$N^{i}$, and $h_{ij}$, the metric tensor induced in the spatial
hypersurfaces $\Sigma_{t}$.
Before the symmetry breaking, the space-time associated
with the rotating star is {\em stationary\/} and
{\em axisymmetric\/}.
We briefly recall the main conclusions for the case where the
star matter is assumed to be constituted by a {\em perfect fluid\/},
the stress energy tensor having the form
\bgeq \label{EQU:TTT}
\vec{T} = (e\!+\!p)\, \vec{u} \otimes \vec{u} + p \, \vec{g} \ .
\edeq
The involved quantities are the fluid proper energy density
$e$, the fluid pressure $p$, the fluid four-velocity $\vec{u}$,
and the space-time metric tensor $\vec{g}$.
Two Killing vector fields $\vec{k}$ and $\vec{m}$ are
linked to the space-time symmetries where $\vec{k}$ is 
time-like at least far from the star and $\vec{l}$ space-like,
its orbits being closed curves. In the case of {\em rigid
rotation\/}, the space-time is {\em circular\/} and the
two-surfaces orthogonal to both $\vec{k}$ and $\vec{m}$
are globally integrable (Carter \cite{CA73}). The coordinates
$t$ and $\phi$ are associated with the both Killing vector
fields $\vec{k}\!=\!\partial_{t}$ and
$\vec{m}\!=\!\partial_{\phi}$ whereas the remaining
coordinates $r$ and $\theta$ can be chosen arbitrarily.
The standard coordinates for stationary axisymmetric systems
are {\em quasi-isotropic\/} coordinates where a {\em
conformally flat\/} metric in the $(r,\theta)$-coordinate
planes is adopted.
The general line element (\ref{EQU:DS2}) specified to these
coordinates reads
\bgea \label{EQU:DSI} g_{\mu\nu} \, {\rm d} x^{\mu} {\rm d} x^{\nu} = -
    N^{2} {\rm d} t^{2} + B^{2} r^{2} \sin^2\theta \,
    ({\rm d} \phi\!-\!N^{\phi} {\rm d} t)^{2} & \\ +
    A^{2} ({\rm d} r^{2}\!+\!r^{2} {\rm d} \theta^{2}) & \ . \nnum
\edea
The shift vector has only one non-vanishing component
$N^{\phi}$, which represents the dragging of inertial
frames by the rotating star.
For this coordinate choice, Bonazzola et al. (\cite{BGSM93}) have
exhibited a compact set of elliptic equations for the
metric potentials
$N$, $N^{\phi}$ and $\tilde{A}$, $\tilde{B}$, the latter
being defined by $\tilde{A}\!=\!A N$, $\tilde{B}\!=\!B N$.
Let us also remind that {\em quasi-isotropic\/}
coordinates satisfy both the {\em minimal distortion\/}
coordinate condition, introduced by Smarr \& York (\cite{SY78}),
and the {\em maximal slicing\/} condition $K\!\equiv\!0$,
where $K$ is the scalar of extrinsic curvature.
When the star deviates from axisymmetry, it is no more
stationary either, as gravitational radiation carries
away energy and angular momentum. However, at the very
beginning of the symmetry breaking, this deviation is
very small, and, consequently, the losses due to
gravitational radiation are negligible. Under this
assumption, and for rigid rotation, there exists a
Killing vector field $\vec{l}$ which is proportional to
the fluid velocity $\vec{u}$ (Carter \cite{CA79}),
\bgeq \label{EQU:KVF} \vec{u} = \lambda\,\vec{l} \ ,
\edeq
where $\lambda$ is a strictly positive scalar function.
In the stationary and axisymmetric case, the Killing
vector $\vec{l}$ is given by
\bgeq \label{EQU:KVG} \vec{l} = \vec{k} + \Omega\,\vec{m} \ .
\edeq
The constant $\Omega$ is the angular velocity defined as
$\Omega\!=\!u^{\phi}/u^{t}$. In the non-axisymmetric case,
we assume (1) the existence of a vector field $\vec{k}$
which is time-like at least far from the star, (2) a vector
field $\vec{m}$ which is space-like everywhere, (3) a constant
$\Omega$, such that $\vec{l}$ defined by (\ref{EQU:KVG})
is a Killing vector field and (4) the fluid velocity $\vec{u}$
is proportional to $\vec{l}$. The Killing vector field $\vec{l}$
is associated with the persisting {\em helical\/} symmetry of
the space-time generated by the non-axisymmetric body which
appears still static in the corotating frame.
\subsection{Matter equations}
\label{SUBSEC:ME}
Based on the assumptions made in the previous section, namely
that (1) the star matter is composed of a single constituent
perfect fluid, and (2) the star rotates rigidly, it is possible
to derive a simple first integral of motion following the
procedure outlined in Bonazzola et al. (\cite{BFG96}).
We first introduce the family of Eulerian observers ${\cal{O}}_{0}$,
whose four-velocity coincides with the future directed unit vector
field $\vec{n}$ orthogonal to the space-like hypersurfaces $\Sigma_{t}$.
The Lorentz factor $\chr$ between these local rest observers
${\cal{O}}_{0}$ and the fluid comoving observers ${\cal{O}}_{1}$ is
given by
\bgeq \label{EQU:GAM} \chr = - \vec{n} \cdot \vec{u} \ .
\edeq
With the baryon chemical potential $\mu$ and the mean baryon
mass $m_{\rm B}$, the {\em log-enthalpy\/} $H$ is defined as
\bgeq \label{EQU:ENT} H \equiv \ln \left( \frac{\mu}{m_{\rm B}
    c^{2}} \right) \ ,
\edeq
which is the relativistic generalization of the Newtonian specific
enthalpy $h$. Introducing $\nu\!=\!\ln N$, we recover the first
integral of motion
\bgeq \label{EQU:FIM} H + \nu - \ln {\chr} = {\rm const} \ ,
\edeq
already familiar from the axisymmetric and stationary case. Note,
however, that in the present case all quantities are functions
of $(r,\theta,\psi)$ where $\psi\!\equiv\!\phi-\Omega\,t$ is the
azimuthal angular variable in the corotating frame .
At the Newtonian limit, we have $H \rightarrow h$, $\nu \rightarrow U$,
$- \ln {\chr} \rightarrow - 1/2\,\Omega^{2} \rho^{2}$, and
(\ref{EQU:FIM}) approaches the classical first integral of
motion where $U$ is the Newtonian potential and $\rho$ the
distance from the rotation axis.
\subsection{Field equations}
\label{SUBSEC:FE}
As announced in Sect.~\ref{SEC:INT}, we apply an approximate set of
field equations derived under the assumptions that (1) the helical
symmetry of space-time is conserved after deviation from the
axisymmetric and stationary configuration, and (2) gravitational
radiation is negligible.

In addition, we only retain the dominant non-axisymmetric
contributions in the field equations up to order {\small 1/2-PN},
their leading relativistic order being less or equal than
$\alpha^{3/2}$ where
\bgeq \label{EQU:PN} \alpha = {\rm max} \left[\frac{\nu}{c^{2}},
    \frac{v^{2}}{c^{2}}, \frac{p}{m_{\rm B} n c^{2}} \right]
\edeq
is the post-Newtonian expansion parameter.
At this level of approximation, the lapse function $N(r,\theta,\psi)$
and the shift vector $N^{i}(r,\theta,\psi)$ have to be considered as
three-dimensional quantities.
The components $N^{r}(r,\theta,\psi)$ and $N^{\theta}(r,\theta,\psi)$
are genuinely three-dimensional contributions and are absent 
at the previous approximation level {\small 0-PN}.
Corrections of higher relativistic order to the metric tensor are
included for the diagonal components via the {\em axisymmetric\/}
potentials $\tilde{A}(r,\theta)$ and $\tilde{B}(r,\theta)$, their
sources being essentially dominated by the fluid pressure whereas
the extra-diagonal terms are again generically non-axisymmetric
quantities and hence are neglected. Accordingly, the spatial metric
tensor reads
\bgeq \label{EQU:XBB} (h_{ij}) = \frac{1}{N^{2}}
\left(\begin{array}{ccc}
    \tilde{A}^{2} & 0                    & 0 \\
    0             & \tilde{A}^{2} r^{2}  & 0 \\
    0             & 0                    & \tilde{B}^{2} r^{2} \sin^{2}\theta
\end{array}\right) \ . \edeq
The choice of $N^{-2}$ as conformal factor of the (nearly
flat) conformal three-metric has proven to be particularly
advantageous, as has been exposed by Bonazzola et al. (\cite{BGSM93}).
It isolates the lapse function as the predominant part of the
gravitational fields, which is underlined by considering the weak
field limit of the corresponding space-time line element, given by
\bgea \label{EQU:WF} g_{\mu\nu} \, {\rm d} x^{\mu} {\rm d} x^{\nu} & = &
   - (1 \! - \! 2 \nu) \, {\rm d} t^{2} + (1 \! + \! 2 \nu) \\
   &  & \times ({\rm d} r^{2} + r^{2} {\rm d} \theta^{2} + r^{2} \sin^{2} \!
   \theta \, {\rm d} \phi^{2} ) \ , \nnum
\edea
which conducts to the Newtonian equation for the gravitational potential
$\nu$.

Having specified the space-time line element for the perturbed
neutron star models, we can proceed to derive the governing field
equations.
The temporal evolution of $h_{ij}$ is determined by
\bgeq \label{EQU:DTM} \partial_{t} h_{ij} + N_{i|j} + N_{j|i} +
    2 N K_{ij} = 0 \ ,
\edeq
with the tensor of extrinsic curvature $K_{ij}$. For $h\!=\!|h_{ij}|
|f_{ij}|^{-1}$, the determinant of the spatial metric tensor,
normalized by that of flat space spherical coordinates
$|f_{ij}|\!=\!r^{4} \sin^{2}\theta$, it follows immediately the
relation
\bgeq \label{EQU:DTH} \partial_{t} h + 2\,(N^{l}{}_{|l} + N K)\,h = 0 \ ,
\edeq
where $K\!=\!K^{j}{}_{j}$ denotes the trace of $K_{ij}$.
Equations (\ref{EQU:DTM}) and (\ref{EQU:DTH}) enable us to derive the
evolution equation of the {\em conformal\/} metric tensor
$\tilde{h}_{ij}\!=\!h^{-1/3}\,h_{ij}$
\bgea \label{EQU:DTC} h^{1/3}
    \partial_{t}[h^{-1/3}\,h_{ij}] +
    \left[N_{i|j} + N_{j|i} -
    \sfrac{2}{3} (N^{l}{}_{|l})\,h_{ij}\right] & \\ \nnum
    + 2 N \left[K_{ij}\!-\!\sfrac{1}{3} K h_{ij}\right] = 0 & \ .
\edea
The lapse function $N$, the only three dimensional quantity
involved in the product $h^{-1/3}\,h_{ij}$, cancels out in
this term. Therefore, the temporal derivative of
$\tilde{h}_{ij}$ vanishes identically. If we further impose
the maximal slicing condition $K\!\equiv\!0$, we can
determine $K_{ij}$ from the metric potentials according to
\bgeq \label{EQU:DFK} K_{ij} = - \left[N_{i|j} +
    N_{j|i} - \sfrac{2}{3} (N^{l}{}_{|l})\,h_{ij} \right]/\,2N \ .
\edeq
Furthermore, the maximal slicing condition yields an elliptic
equation for the lapse function $N$
\bgeq \label{EQU:SCN} N^{|l}{}_{|l} - N[4 \pi (E\!+\!S) +
    K_{kl} K^{lk}]=0\ .
\edeq
Here $E$ stands for the total energy density and $S$ for the trace
of the stress tensor $S^{i}{}_{j}$, all of them measured by the
Eulerian observer ${\cal{O}}_{0}$.

Inserting (\ref{EQU:DFK}) into the momentum constraint equation
\bgeq \label{EQU:EMC} K^{ij}{}_{|j}-8 \pi J^{i} = 0
\edeq
leads immediately to the {\em maximal slicing-minimal distortion\/}
shift vector equation
\bgeq \label{EQU:EMD} N^{i|j}{}_{|j} +
    \sfrac{1}{3} (N^{j}{}_{|j})^{|i} +
    R^{i}{}_{j} N^{j} + 2 K^{ij} N_{|j} +
    16 \pi N J^{i}=0\ ,\ 
\edeq
introduced by Smarr \& York (\cite{SY78}) where $J^{i}$ denotes the
momentum density vector. Indeed, the York minimal distortion
gauge condition $[\partial_{t}(h^{1/3}\,h^{ij})]_{|j}\!\equiv\!0$
is trivially fulfilled, since already the interior of the
square brackets equals 0.
Note that any coordinate system, whose conformal metric
tensor is time-independent, automatically satisfies the minimal
distortion gauge condition. This is notably the case for
isotropic coordinates, but also for our choice of
quasi-isotropic coordinates where the conformal metric is
{\em not\/} that of flat space like in the Wilson scheme, but
time-independent as well.
As a consequence, our approximation of keeping the original
form of the spatial metric except for the lapse function $N$,
being treated as a three-dimensional quantity now, ensures
the coordinates to remain maximal slicing-minimal distortion
coordinates in the three dimensional case after deviation from
the initial stationary and axisymmetric configuration.

The explicit field equations for our particular choice
(\ref{EQU:XBB}) of the spatial metric tensor are then derived
after introduction of the auxiliary variables
\bgeq \label{EQU:VAUX} \tilde{\alpha} \equiv \ln \tilde{A} \ ,
    \quad \tilde{\beta} \equiv \ln \tilde{B} \ , \quad {\rm and}
    \quad \tilde{G} \equiv \tilde{B} \, r\sin\theta \ ,
\edeq
following the procedure outlined in Bonazzola et al. (\cite{BGSM93}).
We obtain the following elliptic equation for the
logarithm $\nu$ of the lapse function $N$
\bgea \label{EQU:SCX} \Delta_{3} \, \nu & = & 4 \pi
    \frac{\tilde{A}^2}{N^2} (E\!+\!S) \\
    \nnum & & +
    \frac{\tilde{B}^2}{2 N^{4}} \, r^{2}\!\sin^{2}\!\theta \,
    (\partial N^{\phi})^{2} -
    \partial \nu \, \partial \tilde{\beta} \ ,
\edea
where $\Delta_{3}$ denotes the three-dimensional flat space
scalar Laplacian with respect to the coordinates $(r,\theta,\psi)$
of the corotating frame
\bgea \label{EQU:LS3} \Delta_{3} & \equiv & \dpar{}{r} +
    \frac{2}{r} \spar{}{r} +
    \frac{1}{r^{2}} \dpar{}{\theta} +
    \frac{1}{r^{2} \tan\theta} \spar{}{\theta} \\
    \nnum & & + 
    \frac{1}{r^{2}\sin^{2}\theta} \dpar{}{\psi} \ ,
\edea
and where the following abridged notation
\bgeq \label{EQU:DEL} \partial \nu \, \partial \tilde{\beta} =
    \spar{\nu}{r} \spar{\tilde{\beta}}{r} +
    \frac{1}{r^{2}} \spar{\nu}{\theta}
    \spar{\tilde{\beta}}{\theta}
\edeq
is used. We define the {\em pseudo-physical\/} components of
the shift vector $N^{i}$ via the following relations
\bgeq \label{EQU:SVPP}
    N^{\tilde{r}}      \equiv N^{r}                     \ , \quad
    N^{\tilde{\theta}} \equiv N^{\theta} \, r           \ , \quad
    N^{\tilde{\phi}}   \equiv N^{\phi}   \, r\sin\theta \ .
\edeq
The shift vector equation (\ref{EQU:EMD}) associated with this
particular frame yields
\bgeq \label{EQU:EMP} \Delta N^{\tilde{p}} +
    \sfrac{1}{3} \, \nabla^{\tilde{p}}(\nabla_{\tilde{q}}
    N^{\tilde{q}}) = S^{\tilde{p}} \ ,
\edeq
with $S^{\tilde{p}}$ specified by (\ref{EQU:SRV}) below.
We further introduce the explicit expressions of the involved
derivative operators, associated with the standard orthonormal
frame of flat space spherical coordinates.
The pseudo-physical components of the three-dimensional flat
space vector Laplacian $\Delta \vec{V}$ are specified as
\bgea \label{EQU:LPVR} (\Delta
   \vec{V})^{\tilde{r}} & \equiv &
   \Delta_{3} V^{\tilde{r}} -
   \frac{2 V^{\tilde{r}}}{r^2} -
   \frac{2}{r^2} \left(\spar{}{\theta} +
   \frac{1}{\tan\theta}\right) V^{\tilde{\theta}} \\
   \nnum & & -
   \frac{2}{r^2\sin\theta} \, \spar{V^{\tilde{\phi}}}{\psi} \ ,
\edea
\bgea \label{EQU:LPVT} (\Delta
    \vec{V})^{\tilde{\theta}} & \equiv &
    \Delta_{3} V^{\tilde{\theta}} -
    \frac{V^{\tilde{\theta}}}{r^2\sin^2\theta} +
    \frac{2}{r^2} \, \spar{V^{\tilde{r}}}{\theta} \\
    \nnum & & -
    \frac{2}{r^2\sin\theta\tan\theta} \, \spar{V^{\tilde{\phi}}}{\psi} \ ,
\edea
\bgea \label{EQU:LPVP} (\Delta
    \vec{V})^{\tilde{\phi}} & \equiv &
    \Delta_{3} V^{\tilde{\phi}} -
    \frac{V^{\tilde{\phi}}}{r^2\sin^2\theta} +
    \frac{2}{r^2\sin\theta} \, \spar{V_{\tilde{r}}}{\psi} \\
    \nnum & & +
    \frac{2}{r^2\sin\theta\tan\theta} \, \spar{V^{\tilde{\theta}}}{\psi} \ .
\edea
We further need to compute the covariant divergence $\nabla_{\tilde{q}}
V^{\tilde{q}}$ which reads
\bgeq \label{EQU:DVS} \nabla_{\tilde{q}} V^{\tilde{q}} \equiv
    \spar{V^{\tilde{r}}}{r} +
    \frac{2 V^{\tilde{r}}}{r} +
    \frac{1}{r} \, \spar{V^{\tilde{\theta}}}{\theta} +
    \frac{V^{\tilde{\theta}}}{r\tan\theta} +
    \frac{1}{r\sin\theta} \, \spar{V^{\tilde{\phi}}}{\psi} \ .
\edeq
Finally, the gradient of a scalar potential $U$ is computed according to
\bgeq \label{EQU:GRAD}
    \nabla^{\tilde{r}}      U \equiv
    \spar{U}{r} \ , \quad
    \nabla^{\tilde{\theta}} U \equiv
    \frac{1}{r} \, \spar{U}{\theta} \ , \quad
    \nabla^{\tilde{\phi}}   U \equiv
    \frac{1}{r\sin\theta} \, \spar{U}{\psi} \ .
\edeq
The pseudo-physical components $S^{\tilde{p}}$ of the actual source,
computed by means of (\ref{EQU:EMD}), read
\bgea \label{EQU:SRV} S^{\tilde{r}} & = & - 16 \pi N J_{r}\ , \quad
    S^{\tilde{\theta}} = - 16 \pi N \frac{J_{\theta}}{r} \ ,
    \ {\rm and} \\
    \nnum
    S^{\tilde{\phi}} & = & - 16 \pi \frac{N \tilde{A}^{2}}{\tilde{B}^{2}}
    \frac{J_{\phi}}{r\sin\theta} - r \sin \theta \,
    \partial N^{\phi} \partial (3 \tilde{\beta} \! - \! 4 \nu ) \ .
\edea
Equations (\ref{EQU:SCX}) and (\ref{EQU:EMP}) together with
(\ref{EQU:SRV}) constitute the 3D-part of our field equations.
The remaining gravitational potentials $\tilde{A}$ and $\tilde{B}$
are computed by means of the dynamical Einstein and the Hamiltonian
constraint equations after integration over $\psi$, which conducts to
the original equations derived in Bonazzola et al. (\cite{BGSM93}).
They are genuine 2D-equations, intimately related to the axisymmetry
and stationarity of the initial configuration (see Gourgoulhon \&
Bonazzola (\cite{GB93}) for a geometrically motivated derivation of the
(3+1)-equations for this case).
For the potentials $\tilde{\alpha}$ and $\tilde{G}$ we then have
\bgeq \label{EQU:E2D1} \Delta_{2} \, \tilde{\alpha} = \left \langle
    8 \pi \frac{\tilde{A}^{2}}{N^{2}} S^{\phi}{}_{\phi} +
    \frac{3\tilde{B}^2}{4 N^{4}} \, r^{2}\!\sin^{2}\!\theta \,
    (\partial N^{\phi})^{2} -
    (\partial \nu)^{2} \right \rangle \ ,
\edeq
\bgeq \label{EQU:E2D2} \Delta_{2} \, \tilde{G} = \left \langle
    8 \pi \frac{\tilde{A}^{2} \tilde{B}}{N^{2}} \, r\sin\theta \,
    (S^{r}{}_{r} \! + \! S^{\theta}{}_{\theta}) \right \rangle \ ,
\edeq
where $\Delta_{2}$ stands for the two-dimensional flat space scalar
Laplacian
\bgeq \label{EQU:LS2D} \Delta_{2} \equiv \dpar{}{r} +
    \frac{1}{r} \spar{}{r} +
    \frac{1}{r^{2}} \dpar{}{\theta} \ ,
\edeq
and where $\langle A \rangle$ denotes the average of $A$ with respect
to the angular variable $\psi$. This ensures the consistency of the
actual sources of (\ref{EQU:E2D1}) and (\ref{EQU:E2D2}) with the
axisymmetry of $\tilde{\alpha}$ and $\tilde{G}$.

To complete the analytic description, we add the matter related
quantities for a perfect fluid whose stress-energy tensor has been
defined by (\ref{EQU:TTT}), expressed in terms of variables of the
(3+1)-formalism. With the Lorentz factor $\chr$, defined by
(\ref{EQU:GAM}), the {\em ``physical''\/} fluid velocity $U_{\hat{a}}$
with respect to the Eulerian observer ${\cal{O}}_{0}$ along the
$a$-th coordinate line is given by
\bgeq \label{EQU:FU} U_{\hat{a}} = \frac{1}{\chr} \, \vec{e}_{\hat{a}} \cdot
    \vec{u} \ ,
\edeq
where $\vec{e}_{\hat{a}}$ is the corresponding spatial unit vector.
For our coordinates, the components $U_{\hat{a}}$ then read
\bgea \label{EQU:UC}
   U_{\hat{r}}      & = & - \frac{\tilde{A}}{N^{2}} \, N^{r} \ , \quad
   U_{\hat{\theta}}   =   - \frac{\tilde{A}}{N^{2}} \, r \, N^{\theta} \ ,
   \ {\rm and} \\
   \nnum
   U_{\hat{\phi}}   & = &   \frac{\tilde{B}}{N^{2}} \, r \sin\theta
   \, (\Omega \! - \! N^{\phi}) \ ,
\edea
and the normalization condition $\vec{u}\!\cdot\!\vec{u}\!=\!-1$ on
the fluid four-velocity $\vec{u}$ yields
\bgeq \label{EQU:GG0} {\chr} = (1 - \vec{U}^{2})^{-1/2} \ .
\edeq
The matter related variables $\chr$, $E$, $J_{i}$ and $S^{i}{}_{j}$,
specified to our coordinate system, take the approximate form
\bgeq \label{EQU:GG1} {\chr} = (1 - U_{\hat{\phi}}^{2})^{-1/2} \ ,
\edeq
\bgeq \label{EQU:SRE} E = {\chr}^{2} ( e \! + \! p ) - p \ ,
\edeq
\bgeq \label{EQU:SRJ} J_\phi = (E \! + \! p)
    \frac{\tilde{B}^{2}}{N^{3}} \, r^2\!\sin^2\!\theta \,
    (\Omega\!-\!N^\phi) \ ,
\edeq
\bgeq \label{EQU:SRS} S^{r}{}_{r} = p \ , \quad
    S^{\theta}{}_{\theta} = p \ , \quad
    S^{\phi  }{}_{\phi  } = ( E \! + \! p ) \, U_{\hat{\phi}}^{2} + p \ ,
\edeq
whereas the remaining components equal 0. By combining (\ref{EQU:SRV})
and (\ref{EQU:SRJ}) one obtains the final form of $S^{\tilde{p}}$,
\bgea \label{EQU:SRW} S^{\tilde{r}} & = & 0 \ , \quad
    S^{\tilde{\theta}} = 0 \ , \ {\rm and} \\
    \nnum
    S^{\tilde{\phi}} & = &
    16 \pi (E\!+\!p)
    \frac{\tilde{A}^{2}}{N^{2}} \, r\sin\theta \,
    (\Omega\!-\!N^\phi) \\
    \nnum & & -
    r \sin \theta \, \partial N^{\phi} \partial (3 \tilde{\beta}
    \! - \! 4 \nu ) \ ,
\edea
where the latter is exactly the same expression as the one presented
in Bonazzola et al. (\cite{BGSM93}) except that the lapse function $N$,
the shift vector component $N^{\phi}$ and the matter term $(E\!+\!p)$
are allowed to depend on $\psi$ now.

Let us finally mention that our analytic scheme yields the {\em exact\/}
solution to the general field and matter equations for two limiting
cases: (1) at the Newtonian limit for an arbitrary deviation
from axisymmetry, and (2) in the axisymmetric case up to arbitrary
relativistic order.
\subsection{Stability of an axisymmetric configuration}
\label{SUBSEC:STAX}
At this point, we can summarize our analytical approach. The elliptic
field equations (\ref{EQU:SCX}), (\ref{EQU:EMP}), (\ref{EQU:E2D1}),
and (\ref{EQU:E2D2}), completed by the first integral equation
(\ref{EQU:FIM}), fully determine an axisymmetric and stationary
equilibrium model, having specified e.g. the central value of the
log-enthalpy $H_{\rm c}$ and the angular velocity $\Omega$
for some particular equation of state.
The above problem can be formulated as a fixed point problem in
some appropriate functional Banach space. Under reasonable physical
assumptions, the induced mapping $\cal{L}$ is contractive, thus
a unique solution exists and the deviation of the sequence
members from the fixed point is bounded by some decaying exponential
function. We refer to Schaudt \& Pfister (\cite{SP96}) for a recent proof
of this statement, though, at present, restricted to weakly relativistic
configurations such as white dwarfs.
The solution scheme consists in solving the three-dimensional
field and matter equations iteratively where as initial guess
a spherical, static matter distribution with a parabolic density
profile is assumed. 
The gravitational potentials are initially set to their flat space
values.
After a few iterations, rotation is switched on, and the solution
converges to the stationary and axisymmetric configuration
fixed by the model parameters $H_{\rm c}$ and $\Omega$.
A particular approximate solution, obtained from a previous
axisymmetric one, remains axisymmetric. The sequence of
equilibrium models is therefore restricted to the
subspace of axisymmetric and stationary ones which is part
of the full configuration space of three-dimensional
quasi-equilibrium configurations. 

At a certain iteration step $J_{0}$, after convergence is considered
to be sufficient, a small perturbation
\bgeq \label{EQU:DNU} \delta \nu \ \propto \ \epsilon \,
    H_{\rm c} \, r^{2} \sin^{2} \! \theta \, (\cos^{2} \! \psi -
    \sin^{2} \! \psi)
\edeq
is added to $\nu\!=\!\ln N$, which excites the $l\!=\!2$, $m\!=\!\pm 2$
mode. Here $H_{\rm c}$ denotes the central value of the log-enthalpy
and $\epsilon$ is a small constant of order $10^{-8}$.
The three-dimensional gravitational potentials $N$ and $N^{i}$ respond
to this perturbation via the field equations and the matter distribution
via the first integral of motion.
The non-axisymmetry of a particular configuration is conveniently
measured by a parameter $\beta$, introduced by means of the Fourier
expansion of $\nu$ in $\psi$,
\bgeq \label{EQU:FTN} \nu \left( r_{\rm e}, \pi/2, \psi 
    \right) = \nu \left ( r_{\rm e}, \pi/2, 0 \right ) +
    \beta \, \cos 2 \, \psi \ ,
\edeq
where $r_{\rm e}$ denotes the mean stellar radius in the equatorial
plane. As mentioned above, $\cal{L}$ applied to {\em axisymmetric\/}
configurations is contractive in some neighbourhood of the previously
constructed axisymmetric solution. Axisymmetric perturbations
will thus decay exponentially. This may be different for the
non-axisymmetric perturbation (\ref{EQU:DNU}), depending on the
influence of the three-dimensional terms in the field and matter
equations.
The stability of the axisymmetric model is decided by inspection of
the behaviour of the non-axisymmetry parameter $\beta$ during the
continued iteration.
Having in mind that we operate in the {\em linear\/} regime, we may
introduce $\kappa$, the {\em amplification factor\/} of the
non-axisymmetry parameter $\beta$ between two successive iterations,
and so the following three cases can be distinguished:
\begin{enumerate}
\item $\kappa < 1$ : $\beta$ decreases exponentially
   and the perturbed configuration converges to
   the non-perturbed axisymmetric configuration ---
   the configuration is {\em secularly stable\/},
\item $\kappa = 1$ : $\beta$ does not change during the
   subsequent iterations --- the configuration is
   {\em se\-cu\-lar\-ly meta-stable\/},
\item $\kappa > 1$ : $\beta$ grows exponentially
   and the perturbed configuration evolves subsequently away
   from the unstable axisymmetric configuration towards a new
   stable triaxial quasi-equilibrium configuration --- the
   configuration is {\em secularly unstable\/}.
\end{enumerate}
To infer the actual stability of a certain configuration,
one has of course to keep in mind the approximate character of
our perturbed equations. In particular, for fully relativistic
configurations, relativistic terms beyond the current
approximation level of order {\small 1/2-PN} which are
not included in the present scheme will possibly alter the
stability against the triaxial secular instability in some a
priori unknown sense.
\section{Numerical code} \label{SEC:NM}
\subsection{Pseudo-spectral method} \label{SEC:PS} 
The numerical implementation of the analytic model described
in Sect.~\ref{SEC:TM} relies on a {\em pseudo-spectral\/}
code which has been derived from the three-dimensional
code presented in Bonazzola et al. (\cite{BFG96}), but enhanced
by the fully three-dimensional treatment of the shift
vector $N^{i}$. The previous code was itself an extension
of an originally axisymmetric code which had been conceived
to construct high precision models of rapidly rotating
stars in general relativity (Bonazzola et al. \cite{BGSM93})
and in the following
applied to study neutron star models based on realistic
equations of state (Salgado et al. \cite{SBGH94}) and to model
neutron stars provided with a strong magnetic field
(Bocquet et al. \cite{BBGN95}). We refer to Bonazzola et al.
(\cite{BGSM93}) for a detailed description of the numerical
techniques which apply essentially to the present three-dimensional
code as well:

The various quantities are expanded in Fourier series in
$\psi$ and $\theta$ and in Chebyshev series in $r$.
The elliptic field equations are solved after expansion
of the actual sources in terms of the angular eigenfunction
bases of spherical harmonics $Y^{m}_{l}(\theta,\psi)$
for $\Delta_{3}$ and $(\cos l\theta, \sin l\theta)$ for
$\Delta_{2}$.
The resulting systems of ordinary differential equations in the
radial variable are conveniently solved in coefficient space.
The elliptic equations are solved exactly in so far as the
computational domain covers the whole space. This allows to
satisfy the exact boundary condition of asymptotic flatness at
spatial infinity and the proper calculation of the source terms
which fill the entire space. In this way, the limitations of
approximate boundary conditions such as a Robin boundary
condition at some finite radius $R$ (York \& Piran \cite{YP82})
can be completely overcome.

Two grids are used to cover the
numerical domain. The inner one embodies the star whereas the
surrounding space is compactified thanks to a change of the
radial variable $u\!=\!r^{-1}$, mapping it onto the finite
exterior grid.

We further want to point out that the present study impressively
demonstrates the ability of a spectral method to exploit
the particular nature of the problem where only a small number
of angular modes in the $\psi$ variable is present. As a
consequence, such low a number as $N_{\psi}\!=\!4$ in the
interval $[\,0,2\pi\,]$ is sufficient to {\em exactly\/} represent
the various quantities up to linear order in $\beta$.
\subsection{Vector Poisson equation} \label{SEC:VP} 
As outlined in Sect.~\ref{SUBSEC:FE}, a major issue for the
computation of triaxial quasi-equilibrium configurations at
the present level of approximation is the calculation of the
shift vector $N^{i}$ by solution of a generalized vector Poisson
type equation of the following form
\bgeq \label{EQU:EMA} \Delta V^{i} + \alpha \, \nabla^{i}(\nabla_{l}
    V^{l}) = S^{i} \ ,
\edeq
where $\alpha$ is a constant number. For $\alpha\!\neq\!-1$,
(\ref{EQU:EMA}) yields an elliptic equation for $V^{i}$,
conducting to a well posed boundary value problem.
The particular case with $\alpha\!=\!1/3$, hence
\bgeq \label{EQU:EMY} \Delta V^{i} +
    \sfrac{1}{3} \, \nabla^{i}(\nabla_{l} V^{l}) = S^{i} \ ,
\edeq
is generic part of the York procedure for the solution of the
initial value problem of general relativity (York \cite{YO83}).
Therefore, numerous attempts of a numerical solution of (\ref{EQU:EMY})
have been made in the past. Bowen (\cite{BO79}) has suggested a
simplified set of equations by solving for the auxiliary variables
$W^{i}$ and $U$ defined through
\bgeq \label{EQU:BOVA} V^{i} = W^{i} - \sfrac{1}{4} \nabla^{i} U \ .
\edeq
With this definition, (\ref{EQU:EMY}) reduces to two Poisson equations
for $W^{i}$ and $U$, namely
\bgeq \label{EQU:BOVE} \Delta W^{i} = S^{i} \ ,
    \quad \Delta U = \nabla_{l} W^{l} \ .
\edeq
This approach has been widely used in analytical and numerical work.
Note, however, that Evans (\cite{EV84}) was able to construct a suitable
Green's function in Cartesian coordinates. It incorporates the
boundary condition $V^{i}\!=\!0$ at spatial infinity and reads
explicitly
\bgeq \label{EQU:EVG} G_{ij} (\vec{x},\vec{x}') = -
    \frac{7}{32 \pi \, | \vec{x} \! - \! \vec{x}' |}
    \left [ \, \delta_{ij} +
    \frac{(x_{i} \! - \! x'_{i}) (x_{j} \! - \! x'_{j})}{7 \, | \vec{x} \! -
    \! \vec{x}' |^{2}} \,
    \right] \ , \
\edeq
where $\delta_{ij}$ is the flat space metric tensor. $G_{ij}$ satisfies
\bgeq \label{EQU:LVG} \left[ \, \Delta G_{ij} +
    \sfrac{1}{3} \, \nabla_{i} \, (\nabla^{l} G_{lj}) \right]
    (\vec{x} \! - \! \vec{x}')=
    \delta_{ij} \, \delta^{(3)} (\vec{x} \! - \! \vec{x}')\ .
\edeq
The elliptic equation (\ref{EQU:EMY}) is hence transformed into the
integral equation
\bgeq \label{EQU:VI} V^{i} (\vec{x}) = \int_{\tens{R}^{3}}
    G^{i}{}_{l} (\vec{x} \! - \! \vec{x}') \, S^{l}(\vec{x}')
    \ {\rm d}^{3}x' \ .
\edeq
In general, calculations have been carried out in Cartesian type
coordinates which considerably simplify the solution of the
elliptic equations (\ref{EQU:BOVE}), because the vector Poisson
equation reduces to {\em independent\/} scalar Poisson equations
for the individual components $V^{x}$, $V^{y}$, and $V^{z}$.
Their use allows further to circumvent the difficulties associated
with the coordinate singularities of spherical or cylindrical
coordinates when using finite difference methods.
Nevertheless, in astrophysical applications the use of spherical
or cylindrical coordinates are often much more adapted to the
geometry of the problem where, for instance, one encounters a
matter distribution with compact support. Note, however, that
because the choice of the local vector basis is arbitrary, one
might employ Cartesian components that are functions of the
spherical coordinates $(r,\theta,\phi)$ as well.
This would allow to adopt Bowen's scheme (\ref{EQU:BOVE}) unaltered.
Boundary conditions may, however, take a simpler form for the
spherical components $V^{r}$, $V^{\theta}$, and $V^{\phi}$.

We propose a new solution method that has been designed to fit
specifically to problems in a spherical coordinate system when
the use of spherical vector components is desirable. Because
our spectral method takes into account the intrinsic regularity
properties of analytic functions, the first step consists in
decoupling the (irrotational) scalar from the (divergence-free)
pure vector part of the involved vector quantities in order to
obtain a system of ordinary Poisson equations analogous to
(\ref{EQU:BOVE}).
We introduce vector fields $\tilde{V}^{i}$ and $\tilde{S}^{i}$,
representing the divergence-free part of $V^{i}$ and $S^{i}$
respectively, as well as two scalar potentials $\Psi$ and $\Phi$.
Appropriate boundary conditions supplied, one has a unique
decomposition
\bgeq \label{EQU:DEC} V^{i} = \tilde{V}^{i} + \nabla^{i} \Psi \ ,
    \quad S^{i} = \tilde{S}^{i} + \nabla^{i} \Phi \ .
\edeq
In a first step, we obtain a Poisson equation for $\Phi$ by
computing the divergence of (\ref{EQU:DEC})
\bgeq \label{EQU:LPH} \Delta \Phi = \nabla_{l} \tilde{S}^{l} \ .
\edeq
Equation (\ref{EQU:EMA}), expressed in terms of the new variables,
then reads
\bgeq \label{EQU:EMV} \Delta \tilde{V}^{i} + \nabla^{i} ((1 \! + \!
    \alpha) \, \Delta \Psi - \Phi) = \tilde{S}^{i} \ .
\edeq
Taking the divergence of (\ref{EQU:EMA}), $H \! = \! ((1 \! + \!
\alpha) \, \Delta \Psi - \Phi)$ turns out to be a harmonic function,
thus satisfying $\Delta H\!=\!0$. We seek regular and bounded
solutions to the initial equation in $\tens{R}^{3}$ and therefore
choose $H\!=\!0$. Consequently, $\Psi$ is determined by the Poisson
equation
\bgeq \label{EQU:LPS} (1 \! + \! \alpha) \, \Delta \Psi = \Phi \ .
\edeq
Combining (\ref{EQU:LPS}) with (\ref{EQU:EMV}), we obtain a vector
Poisson equation for $\tilde{V}^{i}$ which is just (\ref{EQU:EMA})
applied to divergence-free vector fields $\tilde{V}^{i}$ and
$\tilde{S}^{i}$,
\bgeq \label{EQU:LPV} \Delta \tilde{V}^{i} = \tilde{S}^{i} \ .
\edeq
Equations (\ref{EQU:LPS}) and (\ref{EQU:LPV})
resemble (\ref{EQU:BOVE}), but the additional constraint
$\nabla_{l} \tilde{V}^{l}\!=\!0$ considerably simplifies the
solution of (\ref{EQU:LPV}) in spherical coordinates when
the standard orthonormal frame is used. No distinction
between contra- and covariant vector components is made
in this case.
We further drop the tildes on top of the vector indices,
which had been introduced in Sect.~\ref{SUBSEC:FE} to
distinguish the pseudo-physical components.
The explicit form of $\Delta V^{i}\!=\!S^{i}$ then reads
\bgea \label{EQU:LPWR} \Delta_{3} V^{r} & - &
   \frac{2 V^{r}}{r^2} -
   \frac{2}{r^2} \left(\spar{}{\theta} +
   \frac{1}{\tan\theta}\right) V^{\theta} \\
   \nnum & - &
   \frac{2}{r^2\sin\theta} \, \spar{V^{\phi}}{\phi} =
   S^{r} \ ,
\edea
\bgea \label{EQU:LPWT} \Delta_{3} V^{\theta} & - &
    \frac{V^{\theta}}{r^2\sin^2\theta} +
    \frac{2}{r^2} \, \spar{V^{r}}{\theta} \\
    \nnum & - &
    \frac{2}{r^2\sin\theta\tan\theta} \, \spar{V^\phi}{\phi} =
    S^{\theta} \ ,
\edea
\bgea \label{EQU:LPWP} \Delta_{3} V^{\phi} & - &
    \frac{V^{\phi}}{r^2\sin^2\theta} +
    \frac{2}{r^2\sin\theta} \, \spar{V_r}{\phi} \\
    \nnum & + &
    \frac{2}{r^2\sin\theta\tan\theta} \, \spar{V^{\theta}}{\phi} =
    S^{\phi} \ .
\edea
\begin{figure}[t]
\unitlength 1mm
\epsfig{figure=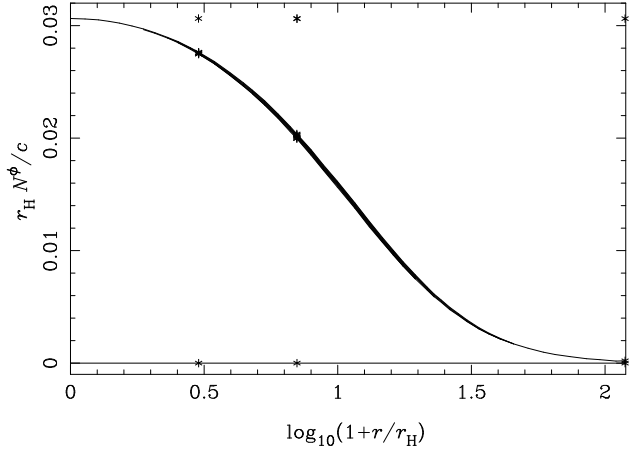,angle=270,width=95mm}
\caption[]{$N^\phi$ for a rapidly rotating Kerr black hole at
$a/M\!=\!0.99$ where $r_{\rm H}$ denotes the radius of the
horizon. The different curves correspond to various values of
$\theta$ ranging from $0^\circ$ to $90^\circ$. The asterisks
indicate the sub-domain boundaries.}
\label{FIG:KNP}
\end{figure}
The solution of (\ref{EQU:LPWR}) to (\ref{EQU:LPWP}) is obtained
in the following way: combining the divergence-free condition on
$V^{i}$,
\bgeq \label{EQU:DVSN} \spar{V^{r}}{r} +
    \frac{2 V^{r}}{r} +
    \frac{1}{r} \, \spar{V^{\theta}}{\theta} +
    \frac{V^{\theta}}{r\tan\theta} +
    \frac{1}{r\sin\theta} \, \spar{V^{\phi}}{\phi} = 0 \ ,
\edeq
with (\ref{EQU:LPWR}) eliminates all components but $V^{r}$ and
leads to an ordinary scalar Poisson equation for the auxiliary
variable $\hat{V}^{r} \! = r \, V^{r}$,
\bgeq \label{EQU:LPVH} \Delta_{3} \hat{V}^{r} = r \, S^{r} \ .
\edeq
In order to simplify (\ref{EQU:LPWT}) and (\ref{EQU:LPWP}) we
express $V^{\theta}$ and $V^{\phi}$ according to
\bgeq \label{EQU:VT} V^{\theta} =
    \frac{1}{r}           \spar{U}{\theta} -
    \frac{1}{r\sin\theta} \spar{W}{\phi} \ , \
    \label{EQU:VP} V^{\phi} =
    \frac{1}{r}           \spar{W}{\theta} +
    \frac{1}{r\sin\theta} \spar{U}{\phi} \ ,
\edeq
where $U$ and $W$ are two auxiliary scalar potentials.
Making once more use of (\ref{EQU:DVSN}), the original equations
(\ref{EQU:LPWT}) and (\ref{EQU:LPWP}) are replaced by
\bgeq \label{EQU:LPU} \frac{1}{r} \spar{}{\theta} \!\!
    \left[ \dpar{U}{r} \! - \! \spar{V^{r}}{r} \right] -
    \frac{1}{r\sin\theta} \spar{}{\phi} \!\!
    \left[ \Delta_{3} W \! - \!
    \frac{2}{r} \spar{W}{r} \right] = S^{\theta} ,
\edeq
\bgeq \label{EQU:LPW} \frac{1}{r\sin\theta} \spar{}{\phi} \!\!
    \left[ \dpar{U}{r} \! - \! \spar{V^{r}}{r} \right] +
    \frac{1}{r}\spar{}{\theta} \!\! \left[ \Delta_{3} W \! - \!
    \frac{2}{r} \spar{W}{r} \right] = S^{\phi} .
\edeq
An equation that only involves $(\partial_{r} U\!-\!V^{r})$
is obtained after differentiation of (\ref{EQU:LPU}),
multiplied by $r\sin\theta$, with respect to $\theta$ and
of (\ref{EQU:LPW}), multiplied by $r$, with respect to $\phi$.
After integration over $r$ where the integration constant is
set to zero in order to make the source term vanish at spatial
infinity, one is left with an equation that only involves the
angular part of the scalar Laplacian, namely
\bgeq \label{EQU:LAW} \frac{1}{r^{2}} \!\!
    \left[ \dpar{}{\theta} +
    \frac{1}{\tan\theta} \spar{}{\theta} +
    \frac{1}{\sin^{2}\theta} \dpar{}{\phi} \right] \!\!
    \left[ \spar{U}{r} \! - \! V^{r} \right] = - S^{r} \ .
\edeq
Since $V^{r}$ has already been determined by means of
(\ref{EQU:LPVH}), $U$ can be computed immediately after
solution of (\ref{EQU:LAW}). In order to fix the lacking
potential $W$, an ordinary integration of $(\ref{EQU:LPW})$
over $\phi$ is carried out, and the integration constant
is set to zero to comply with the required vanishing behaviour
of the source terms of the resulting equation. The final
equation for $\hat{W}^{r}$, defined by $W\!=\!r \, \hat{W}$,
then becomes
\bgeq \label{EQU:LWW} \Delta_{3} \hat{W} = \int_{0}^{\theta}
    \left[ S^{\phi} - \frac{1}{r\sin\theta'}
    \spar{}{\phi} \! \left( \dpar{U}{r} \! - \! \spar{V^{r}}{r}
    \right) \right] {\rm d}\theta' \ .
\edeq
\begin{figure}[t]
\unitlength 1mm
\epsfig{figure=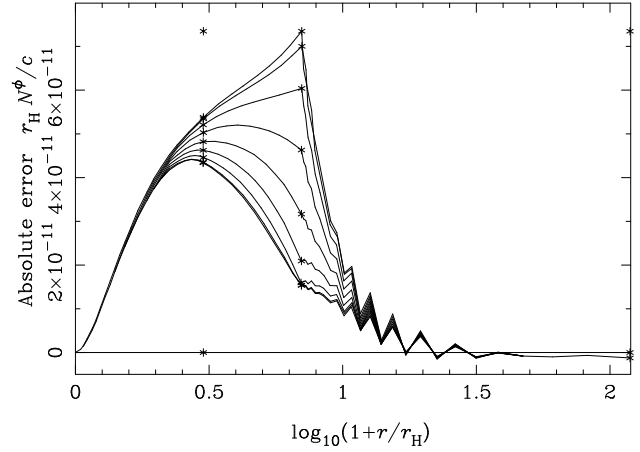,angle=270,width=95mm}
\caption[]{Absolute error in $N^\phi$ corresponding to Fig.~\ref{FIG:KNP}.}
\label{FIG:KNE}
\end{figure}
The system of equations constituted by (\ref{EQU:LPVH}),
(\ref{EQU:LAW}), and (\ref{EQU:LWW}) is equivalent to
(\ref{EQU:LPV}) and significantly simplifies the numerical
solution. Note the absence of $S^{\theta}$ in the source
terms of the final equations. This reflects the dependency of the
components of $S^{i}$ due to the constraint
$\nabla_{l} S^{l}\!=\!0$.

Based on the above approach, we have realized a numerical scheme
that solves (\ref{EQU:EMA}) in a multi domain configuration.
The computational domain may be extended to spatial infinity
thanks to a suitable transformation of the radial variable in
the exterior zone (see Bonazzola et al. (\cite{BGSM93}) for a
previous application of this method).

Numerical tests based on simple analytic functions revealed the well
known {\em evanescent\/} error characteristic for spectral methods
where the numerical errors quickly reach the roundoff limit of
$\simeq\!10^{-14}$ of the employed machine.
It has further been applied to solve the general relativistic field
and matter equations for numerical models of self-gravitating matter
surrounding a rotating black hole which will be presented elsewhere
(Bonazzola \& Gourgoulhon, in preparation).
For vanishing matter density, the Kerr vacuum solution is recovered
and allows a direct comparison with the analytic solution expressed
in quasi-isotropic coordinates.
The {\em coordinate base\/} shift vector component $N^{\phi}$ near the
horizon of a rapidly rotating black hole, characterized by $a/M\!=\!0.99$,
is shown in Fig.~\ref{FIG:KNP} whereas the involved numerical error is
illustrated in Fig.~\ref{FIG:KNE}.
The outside of the black hole is covered by three grids where the
exterior compactified one extends to spatial infinity. The actual grid
resolution is $N_{r}\!=\!33$ in each zone and $N_{\theta}\!=\!7$,
covering the interval $0\!\leq\!\theta\!\leq\!\pi/2$.
The numerical error committed on $N^{\phi}$ nowhere exceeds $10^{-10}$
with respect to the analytic solution and decreases even further to
$\simeq\!10^{-13}$ for higher values of $N_{r}$. Note the very high
precision of the numerical model in spite of the low number of
collocation points, which shows the power of spectral methods for
problems where the involved quantities can be represented by analytic
functions.
\subsection{Validation of the code} \label{SEC:VC} 
The numerical results in the axisymmetric case have been verified
repeatedly by different tests, based on comparison with analytic
solutions or numerical models provided by other authors (Eriguchi
et al., in preparation), as well as by exploiting intrinsic error
indicators like the general relativistic virial identities
(Bonazzola \cite{BO73}; Bonazzola \& Gourgoulhon \cite{BG94};
Gourgoulhon \& Bonazzola \cite{GB94}) or analytic properties of
involved quantities. Typical global errors for rapidly rotating
neutron star models based on a $\gamma\!=\!2$ polytropic equation
of state are about $10^{-6}$.
This value has to be confronted with a global error of $10^{-14}$
for non-rotating, spherical configurations. This difference
stems from a certain deficiency of the spectral approximation of
{\em non-analytic\/} functions.
In the spherical static case, the boundary of the inner spherical
grid can be chosen to coincide with the star surface. As a consequence,
all quantities are analytic in the adjacent sub-domains, and the
numerical error is dominated by the roundoff error of the employed
machine for a moderate number of grid points thanks to the exponential
decrease of the residual numerical errors as a function of the number
of grid points.
For {\em non-analytic\/} functions, the convergence slows down
considerably and is worst for discontinuous functions --- this
behaviour corresponds to the well known Gibbs phenomenon of Fourier
series. In the rotating case, the star surface is located {\em
inside\/} the central grid zone. Therefore, matter variables like
the energy density, whose derivatives
are discontinuous across the star boundary and even become singular
for $\gamma\!>\!2$, are no more analytic functions in this domain
and suffer hence from the above deterioration.
For the current study which is only concerned with a polytropic
equation of state $p\!=\!\kappa\rho^{\gamma}$, this signifies
that $\gamma$ has to be limited to values for which the spectral
approximation still works reasonably well.
In practice, $\gamma$ is bounded by a maximal value
$\gamma_{\rm max}\!=\!3$. In particular, incompressible fluid
configurations such as the Maclaurin or Jacobi spheroids are
beyond the scope of the present code.
Since the range of astrophysical interest is still well within
in the current limits, these restrictions are not prohibitive.

Because of the lack of concurrent studies in the relativistic
regime, the validation of the non-axisymmetric part of the
code is limited to the Newtonian limit where some results
for compressible and incompressible fluids are available,
either obtained by analytic or numerical studies. The different
tests (Bonazzola et al. \cite{BFG96}) include in particular a
confirmation of James' value of the critical adiabatic index,
yielding $\gamma_{\rm crit} = 2.238 \pm 0.002$, which
is in perfect agreement with the predicted value. A classical
result from the theory of Newtonian incompressible fluid bodies
is the particular value $T/|W|_{\rm crit}\!=\!0.1375$ at the
bifurcation point of the secular bar mode instability.
Since we cannot verify this result directly because of the above
restrictions imposed on the equation of state (the incompressible
case corresponds to $\gamma \! \rightarrow \! \infty$), we
studied the dependence of $T/|W|$ for increasing $\gamma$ and found
the successive values to converge to the asymptotic one.
A comparison with results of Ipser \& Managan (\cite{IM81}) and Hachisu
\& Eriguchi (\cite{HE82}) in the compressible polytropic case for a
fixed $\gamma\!=\!3$
showed an agreement of better than 0.5\% for the critical angular
velocity $\Omega_{\rm K}$ and of about 2\% for the values of
$T/|W|_{\rm crit}$ (cf. Table 1 of Bonazzola et al. \cite{BFG96}).

The solution of the 3D-shift vector equation is computed following
the method presented in Sect.~\ref{SEC:VP} and its implementation
is  completely independent from the former scheme where only one
axisymmetric equation for the only non-vanishing shift vector
component $N^{\phi}$ was solved. The deviation of $N^{\phi}$
for identical models is of order $10^{-6}$ and corresponds to the
global numerical error of a particular configuration labeled by the
model parameters $H_{\rm c}$ and $\Omega$.
\section{Results for Polytropes} \label{SEC:RP}
As mentioned in Sect.~\ref{SEC:INT}, earlier investigations in
the Newtonian or at most post-Newtonian regime in general made
use of polytropic equations of state which represent simplified
but within certain limits reasonable and {\em consistent\/}
templates of realistic equations of state.
The {\em critical\/} adiabatic index $\gamma_{\rm crit}$ is defined
as the value of $\gamma$ for which the critical angular
velocity $\Omega_{\rm crit}$ coincides with the Keplerian angular
velocity $\Omega_{\rm K}$, the maximum angular velocity supported
by a rapidly rotating star without mass shedding from the equator.
For $\gamma < \gamma_{\rm crit}$, the star thus cannot rotate
rapidly enough for the secular instability to develop. If
$\gamma > \gamma_{\rm crit}$, there always exists a certain
$\Omega_{\rm crit} < \Omega_{\rm K}$. Provided some mechanism
that fuels the star with rotational kinetic energy, it may be
affected by the secular instability.

\begin{figure}[t]
\unitlength 1mm
\epsfig{figure=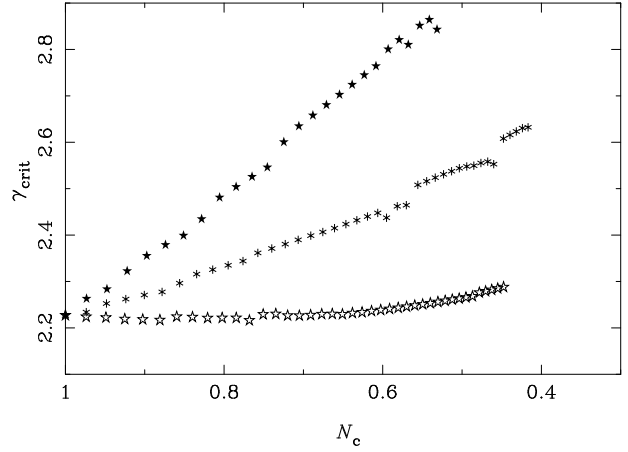,angle=270,width=95mm}
\caption[]{Critical adiabatic index $\gamma_{\rm crit}$ as function
of $N_{\rm c}$, the lapse function at the centre of the star. The upper
curve is obtained with solution of the full 3D-shift vector equation
whereas the lower one illustrates the dependence for the axisymmetric
computation of the shift vector. The intermediate curve shows the
analogous curve for the first case with the axisymmetrized centrifugal
potential.}
\label{FIG:GC}
\end{figure}

Newtonian polytropic stars obey a scaling law and as a
consequence, $\gamma_{\rm crit}$ is a global constant that
does not depend on other model parameters. For relativistic
configurations, effects of general relativity are likely
to influence the bar mode instability.
Therefore, $\gamma_{\rm crit}$ will be a function of the strength
of the relativistic fields which we measure by the value
$N_{\rm c}$ of the lapse function at the centre of the star.
Fig.~\ref{FIG:GC} shows this dependence for the case of the
2D-shift vector equation of our previous investigation as well
as the modified graph for the present approach which includes
the solution of the 3D-shift vector equation.
$\gamma_{\rm crit}$ increases only slightly
in the first case. One even notes a tiny decrease in the weak
field regime. For the present approach, the situation changes
quite dramatically. In the Newtonian regime, both graphs
converge to James' classical value, which, of course, is
required by consistency: for decreasing field strength
the shift vector contribution dies out more rapidly than that
of the lapse function, so that in both cases the asymptotic
space-time line element is given by (\ref{EQU:WF}).
In the relativistic domain, however, the shift vector shows to
inhibit the secular instability more and more efficiently.
The calculations are not continued beyond $\gamma\!=\!3$ where
the steepening gradients of the matter variables severely affect
the spectral approximation.

The question which arises immediately addresses the origin of
the very different behaviour for the both approximations.
Let us recall the classical limit of the first integral of motion
\bgeq \label{EQU:FIDN} h + U - \sfrac{1}{2} \, \Omega^{2}
    \rho^{2} = {\rm const} \ .
\edeq
The evolution of a perturbed configuration is driven by the
mutual influence of the gravitational potential $U$ and
the enthalpy $h$ which are the only three-dimensional quantities
in this case. The centrifugal potential is axisymmetric and
remains unchanged. Since relativistic effects tend to increase
the overall gravitational forces, one might expect that relativistic
corrections to $U$ inhibit the symmetry breaking.
However, the lower graph in Fig.~\ref{FIG:GC},
which illustrates the results of our previous investigation,
shows an only slight growth of $\gamma_{\rm crit}$ and, moreover,
even a {\em destabilizing\/} effect in the weakly relativistic
regime.
In the general relativistic case, (\ref{EQU:FIDN}) is replaced by
\bgeq \label{EQU:FIDR} H + \nu - \ln {\chr} = {\rm const} \ .
\edeq
The crucial point is that now the {\em ``centrifugal''\/} potential
$U_{\Omega}\!=\!-\ln {\chr}$ depends on the azimuthal angle as well.
This means, the angular modulation of $U_{\mathrm{G}}\!=\!\nu$ is
superimposed by that of $U_{\Omega}$, the contribution from the latter
being a purely relativistic effect without Newtonian counterpart.
The variation of $U_{\Omega}$ with respect to $N$ may then enhance
or diminish the triaxial perturbation of the effective relativistic
potential, depending on its sign and, via the first
integral of motion, prevent or favour the triaxial deformation of the
fluid body. The relativistic centrifugal potential $U_{\Omega}$
according to (\ref{EQU:GG1}) reads
\bgeq \label{EQU:UROT} U_{\Omega} = -
    \ln {\chr} = \sfrac{1}{2} \ln (1 \! - \! U_{\hat{\phi}}^{2}) \ ,
\edeq
where $U_{\hat{\phi}}$, the physical fluid velocity in $\phi$ direction
as measured by the local Eulerian observer ${\cal{O}}_{0}$,
specified in (\ref{EQU:UC}), has the form
\bgeq \label{EQU:UROD} U_{\hat{\phi}} = \frac{\tilde{B}}{N^{2}} \,
    r \sin\theta \, (\Omega \! - \! N^{\phi}) \ .
\edeq
In the 2D-shift vector case, the only three-dimensional quantity in
(\ref{EQU:UROD}) is the lapse function $N$. The variation $\delta
U_{\Omega}$ of $U_{\Omega}$
with respect to $N$ adds to $\delta U_{\mathrm{G}}$ with
a {\em positive\/} sign: the triaxial perturbation of the
relativistic potential is enhanced. This explains
the destabilizing effect in the weak field regime before the overall
growth of the relativistic forces related to $U_{\mathrm{G}}$
dominates in the strong field region.
The computation of the shift vector in a three-dimensional fashion
leads to an additional variation of the term $(\Omega \! - \! N^{\phi})$
which contributes to $\delta U_{\Omega}$ with a
{\em negative\/} sign.
It competes with the contribution of the lapse function $N$ to
$\delta U_{\Omega}$, acting in the opposite sense.
It turns out that the combined effect of both contributions is
dominated by the stabilizing effect of the shift vector, which
corresponds to the upper curve in Fig.~\ref{FIG:GC}. From the
Newtonian limit on, the symmetry breaking is increasingly
suppressed. The exclusive influence of the relativistic
gravitational potential $U_{\mathrm{G}}$ can be studied by
averaging $U_{\Omega}$ over $\psi$.
This corresponds to the Newtonian situation, and the resulting
curve for the critical adiabatic index in this case has been
added to Fig.~\ref{FIG:GC}.
It is located between the two other curves, obtained with a
three-dimensional $U_{\Omega}$, and validates
our above reasoning.
We have confirmed numerically that the fluid velocity $U_{\hat{\phi}}$,
appearing in the relativistic centrifugal potential
$U_{\Omega}$,
is the {\em only\/} quantity where the three-dimensional character
of the shift vector actually affects the bar mode instability.
We have further checked the robustness of our results by adding further
three-dimensional terms to the field equations due to an additional
extra-diagonal element in the metric tensor or to treating $\tilde{A}$
and $\tilde{B}$ as {\em ``pseudo''\/} 3D-variables. No noticeable
modification of the results has been observed. These numerical tests
give us confidence in the analytic approximation and indicate that
an additional 3D-treatment of the tensor part of the space-time metric
tensor is unlikely to modify the present results significantly.

As concerns Wilson's approach, we have carried out the above calculations
by imposing $\tilde{A}
\equiv\tilde{B}$ which mimics the ``conformally flat condition'' except
that we do not solve for the fully three-dimensional conformal factor
but rather adopt $\tilde{A}^{2} N^{-2}$ as conformal factor where
$\tilde{A}$ remains axisymmetric.
The global numerical error of the initial axisymmetric configurations
as measured by means of the general relativistic virial theorems is of
order $10^{-4}$ which has to be compared with values of $10^{-6}$
for solution of the exact equations.
The results for the critical adiabatic index $\gamma_{\rm crit}$ appear
to be rather insensitive to this simplification. The actual values
coincide with those by solution of the original equations within a
relative error of about $10^{-3}$ and confirm the validity of
Wilson's simplified scheme for quasi-equilibrium configurations at
least for this particular application.
\section{Conclusion} \label{SEC:CC}
The present work has studied the viscosity driven secular instability
of rapidly rotating stars in general relativity by solution of an
approximate set of field equations related to Wilson's scheme for
quasi-equilibrium configurations. It has been revised to admit a
non-conformal metric tensor which allowed us to recover the exact
equations for stationary and axisymmetric configurations in contrast
with the simplifying ``conformally flat condition''.
The main improvement with respect to our previous study consists
in the fully three-dimensional treatment of the shift vector. 
The numerical scheme has been applied
to configurations built on a polytropic equation of state.
The shift vector shows to strongly enhance the stability of relativistic
configurations compared to the contribution of the lapse function only.
The weak variation of the critical adiabatic index $\gamma_{\rm crit}$
in the previous investigation can be explained by a partial cancellation
of certain non-axisymmetric terms in the first integral of motion.
Comparative calculations adopting the simplifying ``conformally flat
condition'' essentially yield the same results. This is intelligible,
having in mind that even for maximum rotation neutron star models
the deviation of the geometry of curved space from conformal flatness
is of order $10^{-3}$.

Nevertheless, the symmetry breaking is still possible for a mass of
$M_{\rm ns}\!=\!1.4 M_{\odot}$ and an adiabatic index of $\gamma\!=\!2.5$.
A future investigation has to address the question, if realistic models
of neutron star matter still admit the symmetry breaking for
astrophysically relevant masses, as it was the case in our previous
investigation with an axisymmetric shift vector.
\begin{acknowledgements}
J. Frieben gratefully acknowledges financial support by the
{\sc Gottlieb Daimler-und Karl Benz-Stiftung}.
The numerical calculations have been performed on Silicon Graphics
workstations purchased thanks to the support of the SPM department
of the CNRS and the Institut National des Sciences de l'Univers.
\end{acknowledgements}
\end{document}